\documentclass[newstyle,twocolumn,proceedings]{rmaa}
\usepackage{rmaacite}

\renewcommand{\P}[1]{%
\ifnum#1=1\hbox{OW~168--326E}\fi
\ifnum#1=2\hbox{OW~167--317}\fi
\ifnum#1=3\hbox{OW~163--317}\fi
\ifnum#1=5\hbox{OW~158--323}\fi
\ifnum#1=0\hbox{OW~171--334}\fi}

\title{Triggered Star Formation}
\author{J. Palou\v s and S. Ehlerov\' a
  \affil{Astronomical Institute, Academy of Sciences of the Czech Republic,
Prague} }
\fulladdresses{
\item J. Palou\v s and S. Ehlerov\' a: Astronomical Institute, Academy of
Sciences of the Czech Republic, Bo\v cn\' \i \ II 1401, 141 31 Prague 4
Spo\v rilov, Czech Republic (palous@ig.cas.cz, sona@ig.cas.cz).}

\shortauthor{Palou\v s \& Ehlerov\' a}
\shorttitle{Triggered Star Formation}

\keywords{hydrodynamics --- ISM: structure --- ISM:
  supernova remnants --- Galaxies: evolution --- Stars:
  formation}

\abstract{%
The star formation triggered in dense walls of expanding shells will be 
discussed. The fragmentation process is studied using the linear and non-linear
perturbation theory. The influence of the energy input, the ISM distribution
and the speed of sound is examined analytically and by numerical simulations.
We formulate the condition for the gravitational fragmentation of expanding
shells: if the total surface density of the disc is higher than a certain 
critical value, shells are unstable. This value depends on the energy of the 
shell and the sound speed in the ISM.
As an example the formation of OB associations near the Sun will be 
discussed.  We trace their orbits in the Milky Way to see where they have 
been born: 10 - 12 Myr ago  progenitors of 
Scorpius-Centaurus  OB associations and the Orion OB association resided 
together 
within a sheet-like region elongated in the $l = 20^o - 200^o$ direction, 
showing that the local OB associations may be formed as fragments of an   
expanding supershell.   
}

\listofauthors{J.~Palou\v s \& S.~Ehlerov\' a}
\indexauthor{Palou\v s, J.}
\indexauthor{Ehlerov\' a, S.}

\begin{document}

\maketitle

\section{Introduction}
\label{sec:intro}

The gravitational instability in the ISM may develop spontaneously, or it
may be triggered by an external push (Elmegreen; 1998): 
(1) a compression
of pre-existing clouds, (2) an accumulation of gas into a shell that may be
unstable, and (3) cloud collisions. Shell collisions, as considered 
by Chernin et al. (1995), can also influence the star formation. 
Here we focus on the mechanism (2). 

The growth of perturbations on the surface of expanding spherical thin shell
was analyzed in the linear approximation by Elmegreen (1994)
and Vishniac (1994). 
The 
instantaneous maximum growth rate is
\begin{equation}
  \omega = -{3v_{exp} \over R} + \sqrt{{v^2_{exp} \over R^2} +
  \left ({\pi G \Sigma_{sh}
  \over c_{sh}}\right)^2},
  \label{condx}
\end{equation}
where $R$ is the radius of the shell, $v_{exp}$ is its expansion speed,
$\Sigma_{sh}$ is its column density  and $c_{sh}$ is the speed of sound 
within the shell. The  perturbation grows only if $\omega > 0$.

The linear analysis is extended to quadratic terms by 
W\" unsch \& Palou\v s (2001). Quadratic terms give the possibility 
to follow the 
evolution of fragments after the time when the gravitational instability 
starts. Masses of individual fragments and  
their mass spectrum can be evaluated.  

Probably both the spontaneous and triggered star formation operate in 
galaxies and it is difficult to decide, which mechanism is more
important. To discuss  the star formation triggered in expanding shells
without an \` a priori assumption on their shapes, 
we use 3-dimensional simulations. In a numerical code, the condition 
(\ref{condx}) is used and we quantify when and where the expanding 
shell starts to be unstable. 
This approach was first used in Ehlerov\' a et al. (1997), 
here we extend parameter ranges and generalize results.  
We also propose a model of the local system of young stars,
Gould's belt, which may be  the result of triggered of star formation
event in the local ISM.  

\section{Critical surface density} 

To study the influence of the total energy input $E_{tot}$, of the speed of
sound in the ISM $c_{ext}$, of the disk thickness $H$ and of the maximum disk 
density $\rho_0 $ on the gravitational instability of shell,
we use the 3 dimensional numerical model.  
We fix the speed of sound in the
shell $c_{sh}$  and evaluate if the condition for the gravitational 
instability of the shell (\ref{condx}) is fullfilled at some region during the
shell evolution. 

In the parameter space  $\rho_0$ versus $H$ the gravitationally unstable 
and stable regions are separated by surfaces of constant surface density 
$\Sigma_{crit}$. Simulations with different disc profiles
(Gaussian, exponential, multicomponent) show, 
that  $\Sigma _{crit}$ does not depend on it.

There are two types of deviations to this rule:
\begin{itemize}
\item {\bf The blow-out effect.} For thin disks a higher density
$\rho_0 $ than corresponding to 
$\Sigma_{crit}$ is needed for the instability.
The blow-out enables the leakage  of the energy to the galactic halo, leading
to the decrease of the effective energy and pressure pushing the densest parts
of the shell.
\item {\bf The small shell in the thick disk.} 
The low energy shell is generally small, and in  thick disks 
they evolve in an almost homogeneous medium never
reaching dimensions comparable to the thickness of the disk. Consequently, 
the value of the gas surface density of the disk is irrelevant in this case. 
For the instability, the value of $\rho_0 $ has to be higher  than predicted by
$\Sigma_{crit}$ criterion, since a substantial
fraction of the ISM in the disk remains untouched by the shell.
\end{itemize} 

$\Sigma_{crit}$ depends strongly on $c_{ext}$: 
for larger values of $c_{ext}$ the fragmentation
starts at larger values of $\Sigma_{crit}$. 
>From simulations 
with different $c_{ext}$,  $E_{tot}$ we derive the fit:
\begin{eqnarray}
\Sigma_{crit}  = & 0.27 \left ({ E_{tot} \over 10^{51} erg} \right )^{-1.1} 
\nonumber \\
 & \left ({c_{ext} \over km s^{-1}}\right )^{4.1} 
10^{20} cm^{-2}. 
\label{fit}
\end{eqnarray}
With this criterium we can estimate where in the galactic disk the triggered 
star formation operates.

\section{Gould's belt as a triggered star formation event} 

We compute  orbits of individual OB stars from  Gould's belt in the vicinity 
of the Sun backwards in time and analyze
their positions and velocities at different epochs. The volume taken
by members of Scorpius-Centaurus OB association gets smaller going from
now to the past. The smallest volume, 
from which the present OB associations Lower Centaurus-Crux,
Upper Centaurus-Lupus and Upper Scorpius are coming,
is reached between 10 - 12 Myr ago,
when its diameter is less than 100 pc. At that time the 
distance of the center of this region was about 100 pc from the Sun
and all this region was in the first galactic quadrant between 
galactic longitudes $ 10^o < l < 45^o$.
Before that time, even deeper in the past, the orbits of the future 
members of Scorpius-Centaurus OB associations deviate again and the
volume taken by the corresponding test particles restarts to grow.

10 Myr ago the OB associations which form today the Gould's belt, 
particularly the  associations in Orion and Scorpius-Centaurus, have been
closer to each other forming a sheet-like pattern about 500 pc long and less 
than 100 pc wide, with the main axis in the direction $l: 20^o - 200^o$.  

The formation of stars in a region descibed above may be the result of the 
gravitational instability,
fragmentation and subsequent star formation in an expanding shell,
which is deformed by the galactic differential rotation.
The motion of stars formed in a triggered star formation event in 
the expanding shell should be evaluated and compared to the 
observed kinematical parameters of Gould's belt.  

\acknowledgements The authors gratefully acknowledge financial support 
by the Grant Agency of the Academy 
of Sciences of the Czech Republic under the grant No.~A~3003705/1997 and 
support by the 
grant project of the Academy of Sciences of the Czech Republic 
No.~K1048102. JP would like to thank  Guillermo Tenorio-Tagle, Jose Franco 
and organizers of the  ``Ionized Gaseous Nebulae'' meeting in Mexico City for
the financial support during the conference.


\begin{thebibliography}

\bibitem[Chernin<1995>]{Cher95}
  Chernin, A.~D., Efremov, Yu.~N., Voinovich, P.~A., 1995, MNRAS {275}, 313

\bibitem[Ehlerova<1997>]{Sona97}
  Ehlerov\' a, S., Palou\v s, J., Theis, Ch., Hensler, G. 1997, A\&A {328}, 121

\bibitem[Elmegreen<1994>]{BGE94}
  Elmegreen, B.~G. 1994, ApJ {427}, 384

\bibitem[Elmegreen<1998>]{BGE98}
  Elmegreen, B.~G. 1998, in Origins of Galaxies, Stars, Planets and Life,
  ed. C. E. Woodward, H. AQ. Thronson \& M. Shull, ASP Conf. Ser. {148},
  p. 150

\bibitem[Vishniac<1994>]{Vish94}
  Vishniac, E.~T. 1994, ApJ {428}, 186

\bibitem[Wunsch<2001>]{Wun01}
  W\" unsch, R., Palou\v s, J. 2001, A\&A, submitted

\end{thebibliography}
\end{document}